\documentclass[floatfix,showpacs,prl,aps,a4paper,twocolumn]{revtex4}

 \usepackage{bm}
 \usepackage{amsmath}
 \usepackage{amssymb}
 \usepackage{latexsym}
 \usepackage{amsfonts}
 \usepackage{epsfig}
 \usepackage{subfigure}
 \usepackage{color}
 \definecolor{darkblue}{rgb}{0,0,.5}
 \usepackage[linktocpage, colorlinks=true, linkcolor=darkblue, citecolor=darkblue]{hyperref}
 \usepackage[all]{hypcap}

\begin{document}

\title{Dissipation in small systems: A Landau-Zener approach}

\author{Felipe Barra$^1$ and Massimiliano Esposito$^2$}
\affiliation{$^1$ Departamento de Fisica, Falcutad de Ciencias Fisicas y Mathematicas, Universidad de Chile, Santiago Chile \\
$^2$ Complex Systems and Statistical Mechanics, Physics and Materials Science, University of Luxembourg, Luxembourg}

\begin{abstract}
We establish a stochastic thermodynamics for a Fermionic level driven by a time-dependent force and interacting with initially thermalized levels playing the role of a reservoir. The driving induces consecutive avoided crossings between system and reservoir levels described within Landau-Zener theory. We derive the resulting system dynamics and thermodynamics and identify energy, work, heat, entropy and dissipation. Our theory perfectly reproduces the numerically exact quantum work statistics obtained using a two point measurements approach of the total energy and provides an explicit expression for the dissipation in terms of diabatic transitions.
\end{abstract}

\date{\today}

\pacs{
03.65.Yz,   
05.70.Ln,   
05.30.-d    
}

\maketitle

The study of quantum mechanical (QM) diabatic transitions, also called Landau-Zener (LZ) transitions, 
played a major role in many areas of quantum physics since the seminal work of Refs. \cite{Landau32, 
Zener32, Majorana32, Stueckelberg32} (see e.g. the introduction of Ref. \cite{Segal14JCP}). 
They occur between the time-dependent eigenstates of quantum systems driven by time-dependent forces and are often 
interpreted as signature of dissipative processes \cite{Wilkinson88, Wilkinson89, PolkovnikovAP11, PolkovnikovRMP11}.
In this letter, we investigate their connection to dissipation within the framework of quantum 
fluctuation relations such as the quantum Jarzynski relation \cite{EspositoReview, HanggiFTRMP11}. 

In an open system driven by a time-dependent force, the second law states that dissipation or entropy 
production is the sum of the change in the system von Neumann entropy, plus the heat entering the 
reservoir divided by its initial temperature which represents the change in entropy in the reservoir 
if it were ideal (i.e. always at equilibrium). This is shown in Refs. \cite{EspoLindVdBNJP10, 
EspoVdB_EPL_11, ReebWolfNJP14} and in the weak coupling limit (where the reservoir is 
ideal) in Refs. \cite{SpohnLebowitz78, Alicki79, Kosloff13}).
A reversible transformation occurs when entropy production can be neglected.
There are different ways to generate such transformations. One consists in slowly driving an open 
system weakly coupled to an ideal reservoir so that it will remain at any time very close to equilibrium. 
Since heat and entropy change are identical to first order away from equilibrium, entropy 
production is of second order and thus negligible. No notion of QM-adiabaticity enters at this level.  
Another way consists in detaching the system from the reservoir and only consider a driven isolated system 
where the unitary dynamics leaves the system von-Neumann entropy invariant and where no heat is exchanged.
The second law is then empty and the first law trivial (the energy change is the mechanical work done 
by the driving). Once again, QM-adiabaticity plays no role here.

Let us now consider the framework of the quantum Jarzynski relation derived for driven isolated systems 
initially prepared in a canonical equilibrium (using for instance a weak interaction with an ideal 
reservoir which is removed when the driving starts acting) \cite{HTasaki00, Kurchan00}.  
In this case, fluctuations in entropy production are expressed as dissipative work (i.e. the mechanical work minus 
the difference in equilibrium free energy difference corresponding to the final and the initial system Hamiltonian). 
But this dissipation actually only occurs if one reconnects the system to its reservoir after the driving ends. 
This means that in this framework, we are actually dealing with a specific class of driven open 
systems where the driving and the relaxation phase occur separately. The first phase is the driven 
nondissipative dynamics that brings the isolated system to a nonequilibrium state at the expense 
of mechanical work. The second is the nondriven dissipative dynamics starting at the end of the 
first phase when the system is reconnected to the ideal reservoir and ending when 
it has reached equilibrium. The resulting dissipation is the relative entropy distance between 
the nonequilibrium state produced at the end of the first phase and the equilibrium state 
reached at the end of the second one which equals the dissipative work
\cite{VandenBroeck07, JarzynskiEPL09, EspoVdB_EPL_11}.

A special situation occurs if we consider a cyclic driving.
If during the first phase the dynamics is QM-adiabatic, the final state of the system will coincide 
with its initial equilibrium state, and no dissipation will occur during reconnection. 
However, for a cyclic driving generating QM-diabatic transitions, this will not be the 
case and dissipation will ensue. Consequently, for cyclic drivings the 
absence of dissipation is directly linked to QM-adiabaticity. 
For non-cyclic drivings the situation is different. Whether or not the transformation is QM-adiabatic, 
a final nonequilibrium states will generically be produced. In this case the resulting dissipation 
has thus again little to do with the amount of QM-diabatic transitions.

To establish an explicit expression relating dissipation to LZ transitions, we need to go beyond these frameworks
and consider driven systems coupled to reservoirs made of a finite number of levels initially at equilibrium. 
As the driving moves the system levels, they will cross the reservoir levels and a rich dynamics will ensue. 
We will consider the simplest case of a single system level and treat the crossing dynamics within LZ theory. 

\begin{figure}[t]
\begin{center}
\includegraphics[width=.45\textwidth] {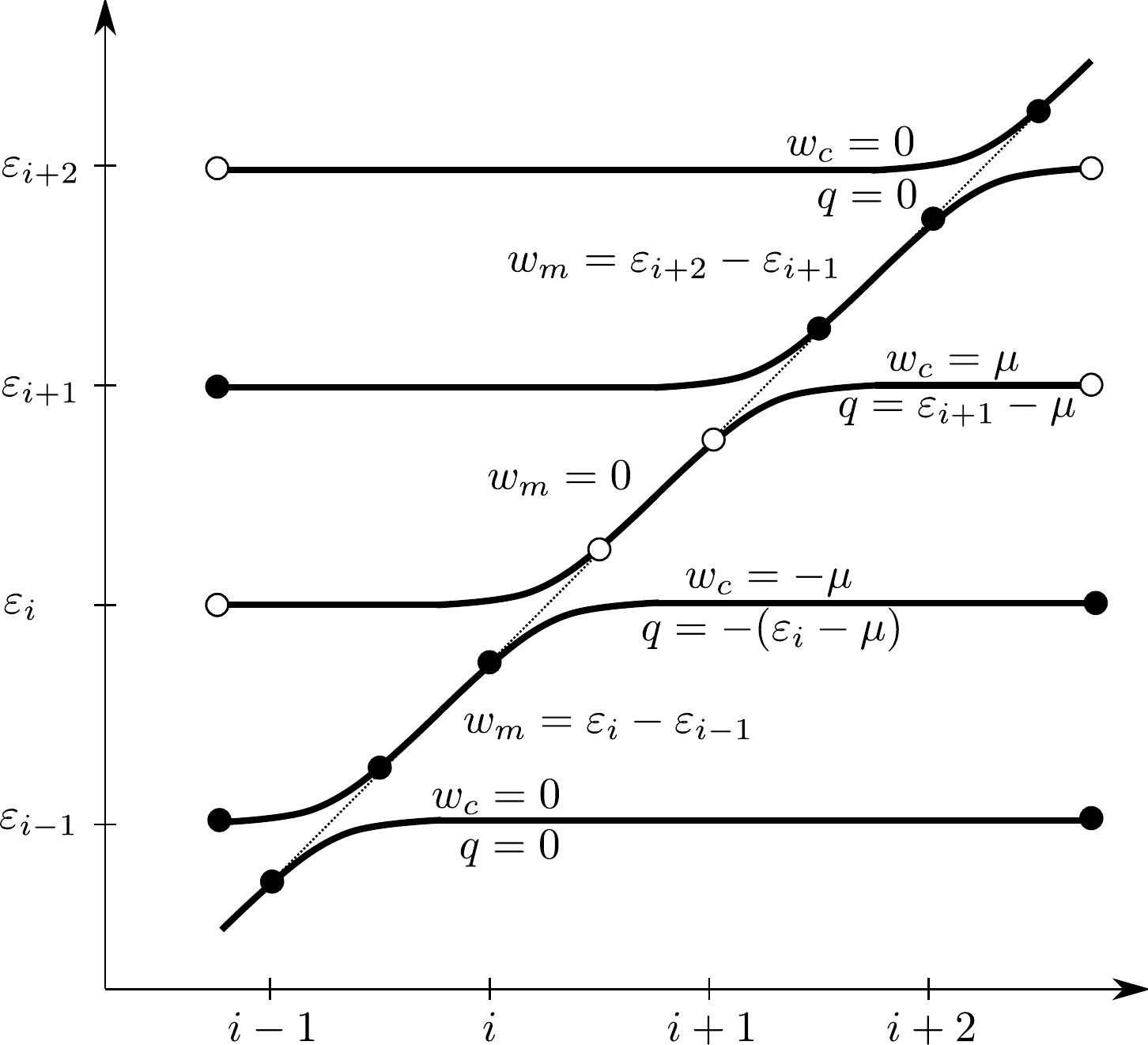}
\caption{Illustration of a possible trajectory when the system level $\epsilon_t$ is driven at 
constant speed $\dot \epsilon$ across equally spaced reservoir levels $\varepsilon_i$. 
Empty (filled) circles denote empty (occupied) levels.}
\label{Draw}
\end{center}
\end{figure}

{\it Model} $-$ 
We assume that we have control over the energy $\epsilon_t$ of a single Fermionic level which constitutes the system.
The reservoir is made of $L$ initially equilibrated Fermionic levels with energy $\varepsilon_i$ ($i=1,\ldots,L$), 
thermal occupation $f_i \equiv f(\varepsilon_i)=1/(e^{\beta(\varepsilon_i-\mu)}+1)$ and spacing 
$\Delta \varepsilon_{i+}=\varepsilon_{i+1}-\varepsilon_{i}$. As usual, $\beta^{-1}=k_B T$ 
and $T$ and $\mu$ are respectively the reservoir temperature and chemical potential.
As the system level is raised, consecutive avoided crossings between the system level and the reservoir levels will occur (see Fig. \ref{Draw}).
The raising speed at the crossing $i$ is denoted $\dot{\epsilon}_i$ and is assumed to remain constant until the next crossing $i+1$ 
(i.e. the raising speed varies slowly between adjacent crossings). The time to go from crossing $i$ to $i+1$ is thus given by 
$\Delta t_{i+}=\Delta \varepsilon_{i+}/\dot\epsilon_i$ and the time at which the crossing with level $i$ occurs is 
$t_i=\sum_{j=1}^{i-1} \Delta t_{j+}$ for $i>1$ and $t_1=0$. The gap between the two levels at an avoided 
crossings $i$ is denoted $\delta_i$ and characterizes the system-reservoir interaction strengths.
It is always assumed smaller than the spacing between the reservoir levels $\Delta \varepsilon_{i+}>\delta_i$, 
so that the system-reservoir dynamics can be treated sequentially (i.e. one reservoir level at the time) and 
within LZ theory \cite{NoriPR2010}. The probability of a QM-diabatic (resp. QM-adiabatic) transition at the 
crossing $i$ is given by $R_i=\exp{\{-\pi \delta_i^2/(2 \hbar \dot{\epsilon}_i)\}}$ (resp. $1-R_i$). 
This probability has been shown to be accurate for times after the crossing longer than 
$t^{\rm lz}_i = \sqrt{\hbar/\dot{\epsilon}_i} \; {\rm max}[1,\sqrt{\delta^2_i/(\hbar \dot{\epsilon}_i)}]$ \cite{SchussPRL89, VitanovPRA99}.
This means that we demand that $\Delta t_{i+} > t^{\rm lz}_i$, which together with $\Delta \varepsilon_{i+}>\delta_i$, 
implies overall that our treatment requires $\Delta \varepsilon_{i+} > \sqrt{\hbar \dot{\epsilon}_i} , \delta_i$.

{\it Dynamics} $-$ 
If $p_{i}$ is the occupation of the system level just before the avoided crossing with $\varepsilon_{i}$, the occupation 
of the system level a time $t^{\rm lz}_i$ after the avoided crossing is given by $p'_{i}=R_i p_{i}+(1-R_i)f(\varepsilon_{i})$.
As the system energy is raised until just before the next crossing at energy 
$\varepsilon_{i+1}$, the probability does not change and thus $p_{i+1}=p'_i$. 
The evolution of the system occupation can therefore be rewritten as a Markov chain with transition 
probabilities at the crossing $i$, $M^-_i=(1-R_i)(1-f_i)$ to empty the system level by filling 
the reservoir one, and $M^+_i=(1-R_i)f_i$ to fill the system level and empty the reservoir one
\begin{equation}
p_{i+1}= (1-M^{-}_i)p_i + M^+_i(1-p_i) . \label{stoch-dyn}
\end{equation}

We proceed with two important remarks. 
First, the state of the reservoir changes as the system level sequentially crosses its levels.
Second, our stochastic model neglects the coherences generated by the quantum dynamics~\cite{NoriPR2010}.
However, our scheme is expected to hold as long as a given reservoir level is not crossed twice. 
In this way, its nonequilibrium state and its coherences resulting from the first interaction will 
not influence the system dynamics anymore.  
\begin{figure}[b]
\begin{center}
\includegraphics[
height=2.0in, width=3.0in ] {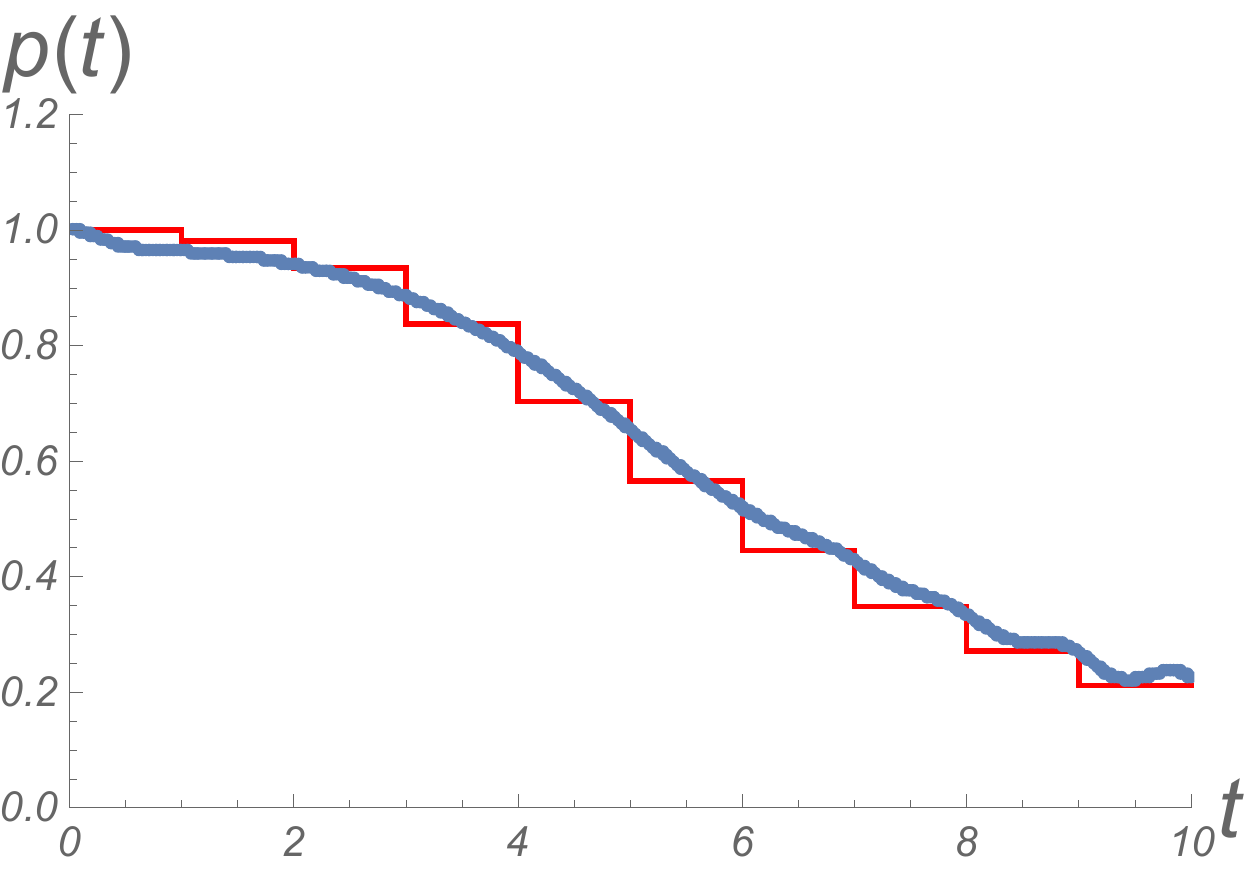}
\caption{System level occupation $p$ for model (\ref{H}) with a protocol $\epsilon_t=\dot\epsilon t=t$.
The system level crosses the $L=9$ reservoir levels with energy $\varepsilon_{i}=i\in\{1,...,9\}$ over a time $t=10$
and $\gamma=0.2$, $\mu=3$, $\beta=1.2$.
The staircase red curve is obtained by solving the stochastic model (\ref{stoch-dyn}) and the continuous blue curve by solving
the numerically exact quantum system for $p=\langle c^\dag c\rangle_t$ with an initial factorized state where the 
reservoir is in the grand-canonical state and the system level is occupied with probability one.}
\label{occup.fig}
\end{center}
\end{figure}
This is confirmed by Fig. \ref{occup.fig}, where we compare the system occupation predicted by our 
stochastic dynamics (\ref{stoch-dyn}) with that predicted using a numerically exact system-reservoir 
quantum mechanical calculation, which is performed as follows. 
The Hamiltonian used is  
\begin{equation}
H(t)=\epsilon_t c^\dag c+\sum_{i=1}^L \varepsilon_i c_i^\dag c_i + \gamma \sum_{i=1}^L (c^\dag c_i+c_i^\dag c),
\label{H}
\end{equation}
where $\epsilon_t=t \dot{\epsilon}$ is the site-energy for the dot, $c^\dag$ and $c$ are its creation and 
destruction operators and $\dot{\epsilon}$ is a constant. The system level and the $\varepsilon_i$ level (with creation 
and destruction operators $c_i^\dag$ and $c_i$) are coupled with a strength $\gamma$ such that the gaps between 
the levels are given by $\delta_i=2\gamma$. Since this is a non-interacting many-body-system, all its properties 
can be obtained from single-body quantities. We thus numerically solved its exact dynamics by mapping the time-dependent 
single-body Schr\"odinger equation into a system of ordinary differential equations, using a Runge-Kutta method (RK4).

We consider two important limiting regimes of the dynamics (\ref{stoch-dyn}).
The {\it QM-adiabatic regime} occurs for slow driving rates $\Delta \varepsilon > \delta \gg \sqrt{\hbar \dot \epsilon}$ 
when no LZ transition occurs because $R_i \to 0$ and as a result $M^+_i=f_i$ and $M^-_i=1-f_i$. 
In this regime, the system instantaneously thermalizes by exchanging its probability with the reservoirs. 
If before the crossing the system is occupied with probability $p_i$ and the reservoir level with probability $f_i$, 
then after the crossing the system occupation becomes $p_{i+1}=f_i$ and the reservoir one $p_i$.
In turn, the {\it QM-diabatic regime} occurs for fast driving rates in terms of LZ theory  
$\Delta \varepsilon > \sqrt{\hbar \dot \epsilon} \gg \delta$. 
In this case a LZ transition always occurs at the crossing because $R_i \to 1$ and as a result $M^+_i=M^-_i=0$. 
The system and the reservoir thus remain unaffected by the crossings, $p_{i+1}=p_i$.

{\it Thermodynamics} $-$ The system average particle number, internal energy and Shannon entropy just 
before the crossing with the reservoir level $i$ is given, respectively, by 
\begin{equation}
 \begin{split}
& \hspace{1cm} N_i = p_{i} \  \ , \  \ E_i = \varepsilon_i p_i \\
& S_i =-k_B p_{i} \ln p_{i}- k_B (1-p_{i})\ln (1-p_{i}).
 \end{split}
 \label{thermquant}
\end{equation}
Across the avoided crossing $i$, the system occupation may change and induce as a 
result a change in particle number, entropy, as well as energy in the form of heat.
In between the crossing with $i$ and $i+1$, the occupation remains unchanged and as 
a result the particle number and the entropy do not change. However, if the level is filled, 
the energy will change under the form of mechanical work (due to changes in the energy level) by an amount $\Delta \varepsilon_{i+}$. 
The average work, $W_{i+} = W_{i+}^{\rm m}+W_{i+}^{\rm c}$, done on the system in going from $i$ to $i+1$ (denoted $i+$ 
in short) thus consists of the mechanical work $W_{i+}^{\rm m}=(\varepsilon_{i+1}-\varepsilon_{i}) p_{i+1}$ 
generated by the driving and the chemical work $W_{i+}^{\rm c}=\mu (p_{i+1}-p_i)$ needed to transfer particles 
from the reservoir to the system.
The corresponding average heat entering the system is $Q_{i+} =(\varepsilon_{i}-\mu) (p_{i+1}-p_{i})$.
In accordance with the first and second law of thermodynamics, the energy and entropy change can be written as
\begin{eqnarray}
&& \Delta E_{i+} = E_{i+1}-E_i= W_{i+}+Q_{i+} \\
&& \Delta S_{i+}= S_{i+1}-S_i= \Sigma_{i+}+Q_{i+}/T .
\end{eqnarray}
Using the {\it local detailed balance} property of LZ rates, $M^+_i/M^-_i=e^{-\beta(\varepsilon_i-\mu)}$, 
the entropy production can be shown to be nonnegative and reads
\begin{eqnarray}\label{EPmodel}
\Sigma_{i+} &=& k_B M^+_{i} (1-p_i) \ln \frac{M^+_{i} (1-p_i)}{M^-_{i} p_{i}}  \\
&&\hspace{0cm}+ k_B M^-_{i} p_i \ln \frac{M^-_{i} p_i}{M^+_{i} (1-p_{i})} - k_B D(p_{i+1} \vert p_{i}) \geq 0 ,\nonumber
\end{eqnarray}
where $D(p \vert p')=p \ln [p/p'] +(1-p)\ln[(1-p)/(1-p')] \geq 0$ denotes the relative entropy.
The detailed calculations for a general stochastic thermodynamics in discrete time are given in the supplementary materials.
Combining the first and second law, the entropy production can also be rewritten as $T \Sigma_{i+}=W_{i+}-\Delta \Omega_{i+}$, 
where $\Delta \Omega_{i+}=\Omega_{i+1}-\Omega_i$ is the change in nonequilibrium grand potential $\Omega_i=E_i-\mu N_i-T S_i$. 
Introducing the dissipated mechanical work $W_{i+}^{\rm diss}=W_{i+}^{\rm m}-\Delta \Omega_{i+}^{\rm eq}$ with the equilibrium grand potential 
$\Omega_{i}^{\rm eq}=k_B T \ln (1-f_i)$, and realizing that $\Omega_{i}-\Omega_{i}^{\rm eq}=k_B T D(p_i \vert f_i)$, we find that entropy 
production can now be expressed as
\begin{equation}\label{DissW}
\Sigma_{i+}=\frac{W_{i+}^{\rm diss}}{T}- k_B D(p_{i+1} \vert f_{i+1}) + k_B D(p_{i} \vert f_{i}) \geq 0.
\end{equation}
The second (resp. third) term on the rhs measures the distance from equilibrium right before crossing $i+1$ (resp. $i$).      

We now consider the thermodynamics of the {\it QM-adiabatic regime}. 
If just before the crossing the system has been prepared in an arbitrary occupation $p_i$, 
using (\ref{EPmodel}) and (\ref{DissW}), we find that the dissipation occurring at the crossing 
is $\Sigma_{i+}=k_B D(p_{i} \vert f_{i})$ and the dissipative work done on the system to lift the 
level from $i$ to $i+1$ is $W_{i+}^{\rm diss}=k_B TD(f_{i} \vert f_{i+1})$.
If the system is initially prepared at equilibrium, $p_1=f_1$ and the first crossing is perfectly reversible $\Sigma_{1+}=0$. 
However, the instantaneous thermalization of the level implies that the subsequent crossings ($i>1$) will be such that 
$p_i=f_{i-1}$ and $p_{i+1}=f_i$ and an amount $\Sigma_{i+}=k_B D(f_{i-1} \vert f_{i})$ will be dissipated at every crossing.
We therefore conclude that in general QM-adiabaticity does not imply thermodynamic reversibility. 
This however becomes true in the limit $\Delta \varepsilon , \delta , \dot \epsilon \to 0$ 
where the inequality $\Delta \varepsilon > \delta \gg \sqrt{\hbar \dot \epsilon}$ is maintained.
Indeed, in this case $\Sigma_{i+}$ as well as $W_{i+}^{\rm diss}$ both tend to zero proportionally to 
order $\Delta \varepsilon^2$ while heat and entropy change become equal, to order $\Delta \varepsilon$.
We now turn to the thermodynamics of the {\it QM-diabatic regime}. 
In this case, the level rises without changes in its initial occupation probability $p_1$. 
As a result no dissipation occurs $\Sigma=0$, and no entropy nor heat is produced. 
Only work is done to lift the energy of the level and $W^{\rm diss}=k_B T\sum_{i} \big( D(p_{1} \vert f_{i+1})- D(p_{1} \vert f_{i}) \big)$.

\begin{figure}[h]
\begin{center}
\includegraphics[height=2.0in, width=3.0in] {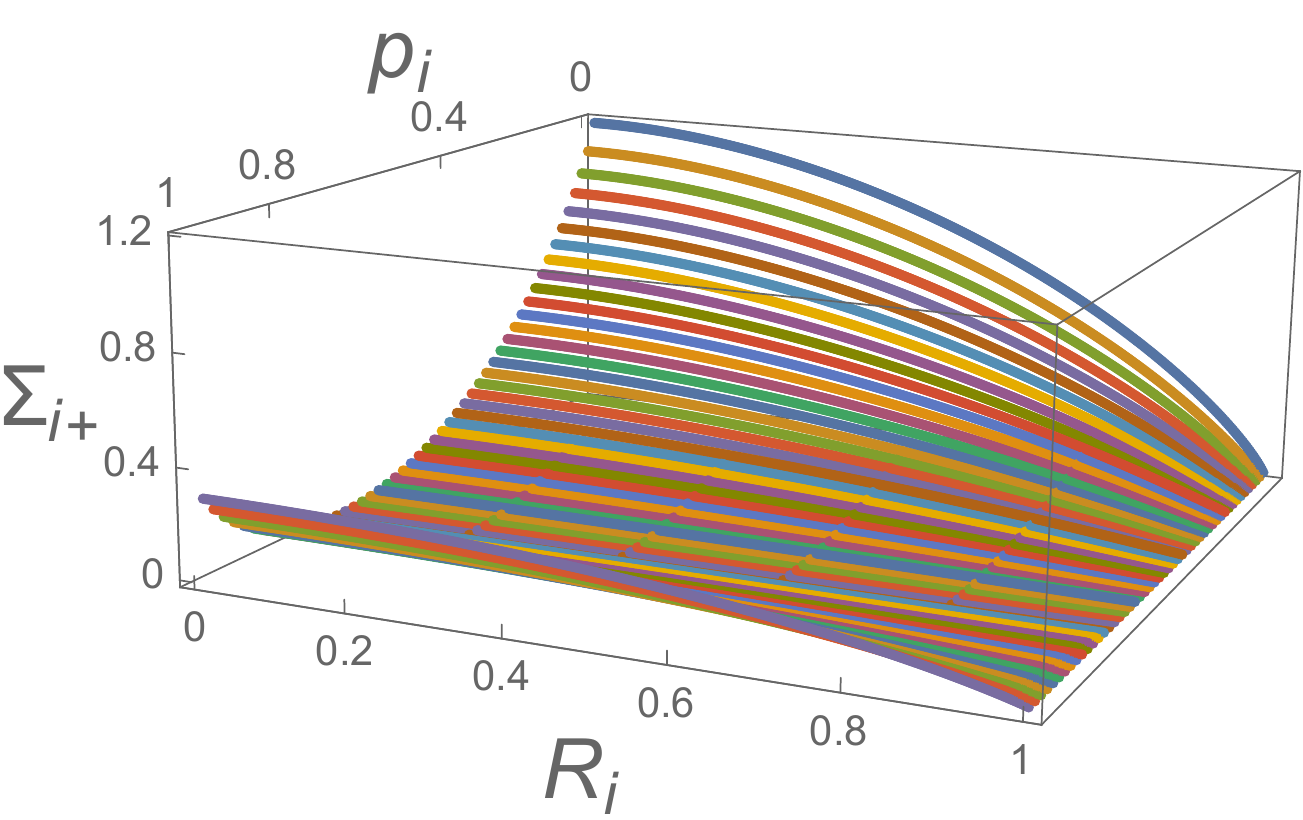}
\caption{Entropy production (\ref{EPmodel}) as a function of the rate of QM-diabatic transitions $R$ 
and of the initial system level occupation $p_{i}$ before the crossing, for $f_i=0.75$.}
\label{EP.fig}
\end{center}
\end{figure}
The dissipation (\ref{EPmodel}) or (\ref{DissW}) across a transition is represented in Fig. \ref{EP.fig} 
as a function of the rate of QM-diabatic transitions $R$ and of the initial system level occupation $p_{i}$ before 
the crossing and for $f_i=0.75$. As announced, the dissipation vanishes when $R \to 1$ (QM-diabatic limit) 
independently of $p_{i}$ or when $R \to 0$ (QM-adiabatic limit) if $p_{i}=f_i$.

\begin{figure}[t]
\begin{center}
\includegraphics[height=2.0in, width=3.0in ] {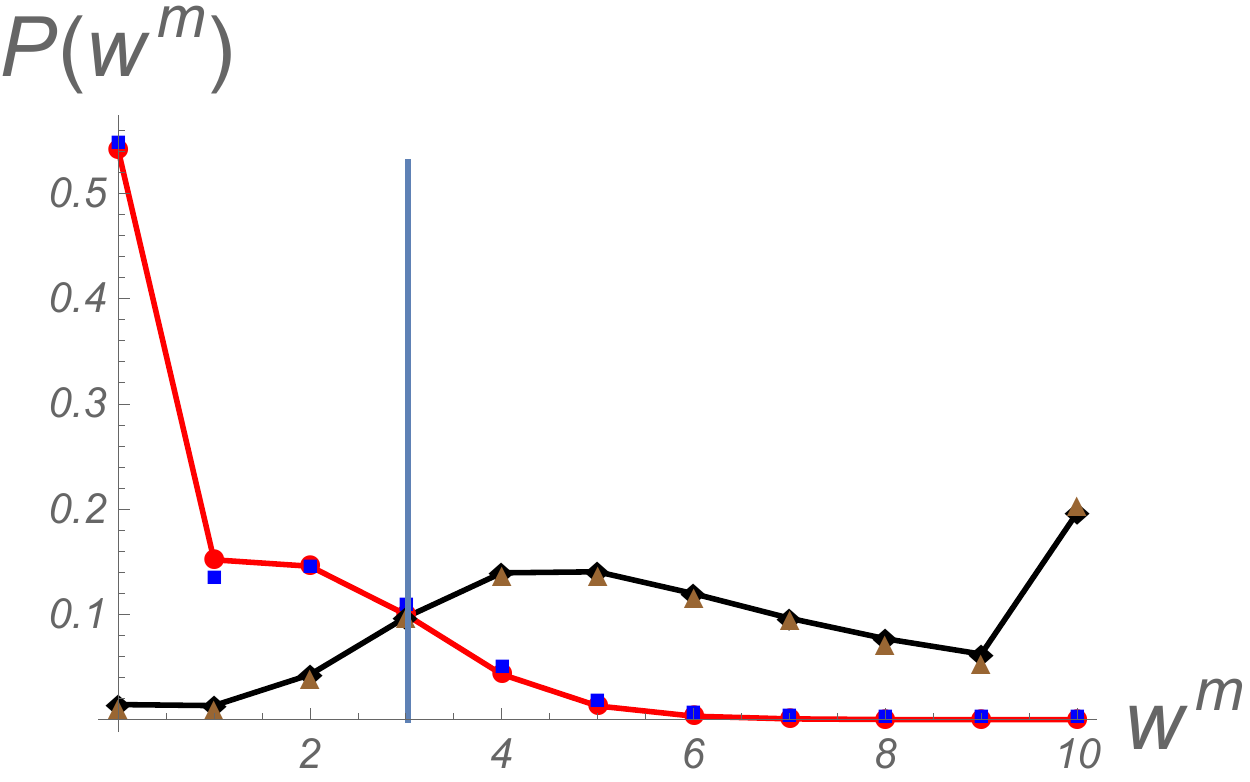}
\caption{Forward $P(w^{\rm m})$ (resp. backward $\tilde{P}(-w^{\rm m})$) work distribution calculated using LZ-theory, 
in black joined (resp. red joined), and using the full quantum dynamics, in brown (resp. blue). 
The vertical line marks the value of $\Delta \Omega^{\rm eq}$. Same parameters as Fig. \ref{occup.fig}.}
\label{workpdf.fig}
\end{center}
\end{figure}

{\it Mechanical work fluctuations} $-$ We can use our stochastic thermodynamic description of the system to 
study mechanical work fluctuations. Thanks to the local detailed balance property of the rates, the work 
fluctuation theorem can be derived following a procedure almost identical to that detailed in 
Refs. \cite{Crooks00, Crooks98, EspoBulImpPRE14} and reads 
$\ln P(w^{\rm m})/\tilde{P}(-w^{\rm m}) = \beta (w^{\rm m}-\Delta \Omega^{\rm eq})$. 
$P(w^{\rm m})$ denotes the probability that the external force performs a mechanical work $w^{\rm m}$ 
when driving the system (initially at equilibrium) according to a given forward protocol. 
$\tilde{P}(w^{\rm m})$ denotes the same probability when the driving protocol is time-reversed and 
the system is initially at equilibrium with respect to the final value of the forward driving protocol. 
In Fig. \ref{workpdf.fig}, these two distributions obtained using our stochastic model are shown to be in 
excellent agreement with those obtained using the numerically exact quantum dynamics in the total system 
with Hamiltonian (\ref{H}).
In this latter case, the mechanical work is obtained from the energy changes resulting from 
a two point projective measurement of the total system energy at the beginning and at the 
end of the process \cite{HTasaki00, Kurchan00, EspositoReview, HanggiFTRMP11}. 

{\it Continuous time limit} $-$ 
We now consider that we operate close to the QM-diabatic regime where $\Delta \varepsilon > \sqrt{\hbar \dot{\epsilon}} > \delta$
and we assume that the coupling $\delta_i$, the driving $\dot{\epsilon}_i$, and the reservoir density of states 
$d=1/\Delta \varepsilon$ vary smoothly with $i$. In this regime, the rate of QM-diabatic transition can be 
expanded as $R \approx 1- \frac{\delta^2}{\hbar \dot{\epsilon}} \frac{\pi}{2}$.
Introducing the reservoir density of states $d_i=1/\Delta \varepsilon_i$ and remembering that 
$\Delta t_{i+}=1/(\dot{\epsilon}_i d_i)$, the dynamics 
$(p_{i+1} - p_i)/\Delta t_{i+} = (f_i-p_i)(1-R_i)/\Delta t_{i+}$ can be treated as a 
continuous time master equation $d_t p=w^+(1-p)-w^{-}p$ with Fermi golden rule rates
\begin{eqnarray}
&&w^+=\frac{\pi \delta^2(\epsilon_t) d(\epsilon_t)}{2 \hbar} f(\epsilon_t) \\
&&w^-=\frac{\pi \delta^2(\epsilon_t) d(\epsilon_t)}{2 \hbar} (1-f(\epsilon_t)) .
\end{eqnarray}
These rates satisfy local detailed balance and thus a consistent stochastic 
thermodynamics ensues \cite{Seifert12Rev, Esposito12, EspVDBRev2014}.
Note that their explicit dependence on the driving speed $\dot{\epsilon}_i$ has disappeared.
The restriction to be close to the QM-diabatic regime however puts the limit of 
reversible transformations outside the realm of validity of this description.  
It is interesting to note that exactly the same dynamics can be derived by assuming that the system 
is weakly coupled to a continuous reservoir in the Born-Markov secular approximation \cite{Breuer02}. 
However, in this case the above restriction does not hold and the limit of reversible transformations 
is reachable. This may indicate that the restriction $\Delta \varepsilon > \delta$ could be 
loosened as also suggested by the results of Ref. \cite{KayanumaJPB85}. 

{\it Conclusions} $-$ 
We established a stochastic thermodynamics for a single system level interacting with a finite set of initially 
thermalized reservoir levels. This enabled us to derive an explicit expression for dissipation in terms of rates 
of QM-diabatic transitions. Thermodynamically reversible transformations require QM-adiabaticity but also 
a vanishing reservoir level spacing as well as a vanishing system-reservoir interacting strength.  
Our treatment can be immediately generalized to multiple system levels as long as the system energies do not cross. 
This latter case with crossings would require some more care. 
We emphasize that our study is very different from other studies which considered the crossing dynamics 
between system levels in contact with continuous reservoirs \cite{RammerAoPRB91, StenholmPRA93, Segal14JCP}.  

{\it Acknowledgments} $-$
This work was made possible by the COST Action MP1209.
F.B. acknowledges support from Fondecyt grant 1151390. 
M.E. is supported by the National Research Fund, Luxembourg (project FNR/A11/02).


\section*{Supplementary Material: Stochastic thermodynamics in discrete time}

The stochastic process considered in the letter is a special case of a Markov chain in discrete time satisfying local detailed balance.
We now derive stochastic thermodynamics for this general case.  
We consider discrete times $i=0,1,2,\dots$ and a finite state space $m=0,1,2,\dots,M$. 
The probability to find the system in state $m$ at time $i$ is denoted $p_{m}(i)$ and evolves according to the Markov chain 
\begin{equation}
p_m(i+1) = \sum_{m'} M_{m m'}(i) p_{m'}(i) \label{MChain},
\end{equation}
where the transition matrix satisfies $\sum_{m} M_{m m'}(i)=1$. 
We further decompose the latter into contributions from different reservoirs $\nu=1,2,\dots,R$: $M_{m m'}(i)=\sum_{\nu} M_{m m'}^{(\nu)}(i)$.
We also introduce the time dependent energy $e_m(i)$ and the number of particles $n_m$ of state $m$
and assume that the transition matrix satisfies {\it local detailed balance} 
\begin{equation}
\ln \frac{M_{m m'}^{(\nu)}(i)}{M_{m'm}^{(\nu)}(i)} = - \frac{\big(e_{m}(i)-e_{m'}(i) \big)-\mu_{\nu}(i)\big(n_m-n_{m'}\big)}{T_{\nu}(i)},
\label{LDB}
\end{equation}
where $T_{\nu}$ and $\mu_{\nu}$ is the temperature and chemical potential of reservoir $\nu$ and $k_B=1$. 
The model considered in our letter has two states: $m=0$ when the level is empty and $m=1$ when it is 
occupied with probability $p_0=1-p$ respectively $p_1=p$.
Furthermore, $n_0=0$, $e_0=0$, $n_1=1$, $e_1(i)=\epsilon(i)$.

The average system energy and number of particles is given by 
\begin{eqnarray}
&&E(i) = \sum_m e_m(i)p_m(i) \\
&&N(i) = \sum_m n_m p_m(i).
\end{eqnarray}
The energy and particle current entering the system from reservoir $\nu$ are
\begin{eqnarray}
&&I_E^{\nu}(i) =  \sum_{m,m'} \big( e_{m}(i)-e_{m'}(i) \big) M_{mm'}^{\nu}(i) p_{m'}(i) \\
&&I_M^{\nu}(i) =  \sum_{m,m'} \big( n_{m}-n_{m'} \big) M_{mm'}^{\nu}(i) p_{m'}(i)
\end{eqnarray}
and the average heat entering the system from reservoir $\nu$ is
\begin{equation}
Q^{\nu}(i) = I_E^{\nu}(i) - \mu_{\nu} I_M^{\nu}(i) .
\end{equation}
The average work done on the system is made of mechanical and chemical work 
\begin{eqnarray}
&&W(i) =  W_{mech}(i) + W_{chem}(i) \\
&&W_{mech}(i) =  \sum_m \big( e_m(i+1)-e_m(i) \big) p_{m}(i+1) \nonumber \\
&&W_{chem}(i) =  \sum_{\nu} \mu_{\nu} I_M^{\nu}(i) \nonumber .
\end{eqnarray}
The first law of thermodynamics ensues and is given by
\begin{equation}
\Delta E(i) = E(i+1) - E(i) = W(i) + \sum_{\nu} Q^{\nu}(i).
\end{equation}
We now define the Shannon entropy of the system
\begin{equation}
S(i)=-\sum_{m} p_{m}(i) \ln p_{m}(i).
\end{equation}
The second law of thermodynamics reads
\begin{equation}
\Delta S(i) \equiv S(i+1) - S(i) = \sum_{\nu} \frac{Q^{(\nu)}(i)}{T_{\nu}} + \Sigma(i) ,
\end{equation}
where the entropy flow is given by
\begin{eqnarray}
\sum_{\nu} \frac{Q^{(\nu)}(i)}{T_{\nu}} = 
- \sum_{\nu,m,m'} M_{mm'}^{(\nu)}(i) p_{m'}(i) \ln \frac{M_{mm'}^{(\nu)}(i)}{M_{m'm}^{(\nu)}(i)} 
\end{eqnarray}
and the entropy production is defined as
\begin{equation}
\Sigma(i) = \sum_{\nu,m,m'} M_{mm'}^{(\nu)}(i) p_{m'}(i) \ln \frac{M_{mm'}^{(\nu)}(i) p_{m'}(i)}{M_{m'm}^{(\nu)}(i) p_{m}(i+1)} .
\end{equation}
This quantity is non-negative as can be shown using the inequality $-\ln X \geq X-1$. Indeed, 
\begin{eqnarray}
&&\hspace{-0.3cm}\Sigma(i) \geq \sum_{\nu,m,m'} M_{mm'}^{(\nu)}(i) p_{m'}(i) 
\big( \frac{M_{m'm}^{(\nu)}(i) p_{m}(i+1)}{M_{mm'}^{(\nu)}(i) p_{m'}(i)} -1 \big) \nonumber \\
&&=\sum_{\nu,m,m'} \big( M_{m'm}^{(\nu)}(i) p_{m}(i+1)- M_{mm'}^{(\nu)}(i) p_{m'}(i)  \big) =0 .\nonumber
\end{eqnarray}
It is zero when detailed balance is satisfied, i.e. when 
\begin{equation}
M_{mm'}^{(\nu)}(i) p_{m'}(i) = M_{m'm}^{(\nu)}(i) p_{m}(i+1).
\end{equation}
The entropy production can also be rewritten as
\begin{eqnarray}
\Sigma(i)&=& \sum_{\nu,m,m'} M_{mm'}^{(\nu)}(i) p_{m'}(i) \ln \frac{M_{mm'}^{(\nu)}(i) p_{m'}(i)}{M_{m'm}^{(\nu)}(i) p_{m}(i)} \nonumber \\
&&- \sum_{m} p_m(i+1) \ln \frac{p_m(i+1)}{p_m(i)}.
\end{eqnarray}
The first term is the non-negative term that survives in the continuous time limit (see e.g. \cite{Gaspard04b}) 
while the second one is negative and vanishes. A similar expression was also found in \cite{Altaner12, AltanerThesis}.

\end{document}